\documentclass[%
 aip,
 amsmath,amssymb,
 reprint,%
]{revtex4-1}

\usepackage{graphicx}
\usepackage{dcolumn}
\usepackage{bm}

\usepackage[utf8]{inputenc}
\usepackage[T1]{fontenc}
\usepackage{siunitx}

\usepackage{color} 
 
\usepackage[linktocpage,colorlinks=true,linkcolor=blue,urlcolor=blue,
citecolor=blue,hyperfigures]{hyperref}

\DeclareRobustCommand*{\citen}[1]{%
  \begingroup
    \romannumeral-`\x 
    \setcitestyle{numbers}%
    \cite{#1}%
  \endgroup
}
\begin{document}

\title[]{Vapor-liquid equilibrium of water with the MB-pol many-body potential }

\author{Maria Carolina Muniz}
\altaffiliation[]{These authors contributed equally to this work.}
\affiliation{Department of Chemical and Biological Engineering, Princeton University, Princeton, New Jersey 08544, USA}
 
\author{Thomas E. Gartner III} 
\altaffiliation[]{These authors contributed equally to this work.}
\affiliation{Department of Chemistry, Princeton University, Princeton, New Jersey 08544, USA}

\author{Marc Riera}
\affiliation{Department of Chemistry and Biochemistry, University of California, San Diego, La Jolla, California 92093, USA}

\author{Christopher Knight}
\affiliation{Computational Science Division, Argonne National Laboratory, Argonne, Illinois 60439, USA}

\author{Shuwen Yue}
\affiliation{Department of Chemical and Biological Engineering, Princeton University, Princeton, New Jersey 08544, USA}

\author{Francesco Paesani}
\email{fpaesani@ucsd.edu}
\affiliation{Department of Chemistry and Biochemistry, University of California, San Diego, La Jolla, California 92093, USA}
\affiliation{Materials Science and Engineering, University of California San Diego, La Jolla, California 92093, USA}
\affiliation{San Diego Supercomputer Center, University of California San Diego, La Jolla, California 92093, USA}

\author{Athanassios Z. Panagiotopoulos}
\email{azp@princeton.edu}
\affiliation{Department of Chemical and Biological Engineering, Princeton University, Princeton, New Jersey 08544, USA}

\date{\today}

\begin{abstract}
 Among the many existing molecular models of water, the MB-pol many-body potential has emerged as a remarkably accurate model, capable of reproducing thermodynamic, structural, and dynamic properties across water's solid, liquid, and vapor phases. In this work, we assessed the performance of MB-pol with respect to an important set of properties related to vapor-liquid coexistence and interfacial behavior. Through direct coexistence classical molecular dynamics simulations at temperatures 400 K < $T$ < 600 K, we calculated properties such as equilibrium coexistence densities, vapor-liquid interfacial tension, vapor pressure, and enthalpy of vaporization, and compared the MB-pol results to experimental data. We also compared rigid vs. fully flexible variants of the MB-pol model and evaluated system size effects for the properties studied. We found that the MB-pol model predictions are in good agreement with experimental data, even for temperatures approaching the vapor-liquid critical point; this agreement was largely insensitive to system size or the rigid vs. flexible treatment of the intramolecular degrees of freedom. These results attest to the chemical accuracy of MB-pol and its high degree of transferability, thus enabling MB-pol's application across a large swath of water's phase diagram.     
 
\end{abstract}

\maketitle

\section{Introduction}

\par Water has been the focus of many computational studies\cite{ Gallo2016, Cisneros2016a, Vega2011, Gillan2016}, given its biological and industrial importance and its often intriguing behavior. However, developing a model that accurately describes the thermophysical properties of water from gas to the condensed phases is a challenging task due to the complexity of water's intermolecular interactions. As a representative example, one important set of phenomena that has been the focus of many computational studies is water's vapor-liquid equilibrium (VLE) \cite{Chen2000,Moucka2013a, Ismail2006, Sakamaki2011, Sega2017, Wohlfahrt2020}. Accurately predicting water's VLE behavior is important due to the key role it plays in several industrial applications, such as distillation and the production of electricity via steam turbines. The early stages of the liquid-vapor phase transformation, marked by bubbles appearing in liquid water in a process termed cavitation, is also a concern to engineering applications, since it often results in damages to pumps and turbines.   

\par In an ideal case, correlated quantum chemistry methods would be used to sample water's potential energy surface (PES) from first principles, but the computational expense of such techniques make comprehensive predictions of water's bulk thermodynamic properties and phase behavior largely out of reach with current computational resources. Due to the high computational cost involved in \textit{ab initio} molecular dynamics (MD) simulations, and despite broad interest and the potential utility of predictive simulations of water VLE, relatively few `first principles' investigations of water VLE have been reported to date \cite{McGrath2006, McGrath2006_JPC, Baer2011, Wohlfahrt2020}.  


\par As an alternative to costly quantum chemistry methods, empirical force fields have been developed to study water-related phenomena via classical molecular simulations \cite{Tainter2015, Kumar2008, Berendsen1987, Jorgensen1983, Horn2004, Abascal2005, Paricaud2005, Lobanova2015, Jiang2016, Kiss2013, Ren2003, Ponder2010, Wang2013, Laury2015}. To date, the most-used empirical representations of water consist of fixed-charge, rigid, non-polarizable models built on pairwise interaction potentials, such as SPC/E\cite{Berendsen1987}, TIP3P\cite{Jorgensen1983} and TIP4P/2005\cite{Abascal2005}. Due to their simplicity and computational efficiency, these models allow for simulations of significantly larger systems and/or greater timescales than \textit{ab initio} approaches and are often able to reproduce many important properties of liquid water \cite{Vega2011}. However, when it comes to reproducing simultaneously vapor phase and VLE properties, such as vapor pressures and densities, surface tension, and critical properties,  many pairwise rigid and/or non-polarizable models fall short \cite{Vega2006, Sakamaki2011, Vega2011}, often due to their inability to represent polarization and/or many-body effects. 

\par With that in mind, more advanced empirical models have been developed, such as E3B \cite{Tainter2015, Kumar2008}, which includes a three-body term, and polarizable models, such as AMOEBA\cite{Ren2003, Ponder2010, Wang2013, Laury2015}, the Baranyai–Kiss models (BK3) \cite{Kiss2013} and the Hydrogen-Bonding Polarizable (HBP) model \cite{Jiang2016}. The inclusion of polarization results in better accuracy when computing properties related to VLE and the gas phase, such as virial coefficients, vapor pressures, and critical properties \cite{Jiang2016, Kiss2013, Ren2003, Ponder2010, Wang2013, Laury2015}. However, empirical models parameterized to reproduce certain experimental results may still lack in transferability, which could limit their use to predict water's properties across the phase diagram. Furthermore, studies suggest that including many-body polarization effects might be a way of improving transferability in empirical models, especially for calculating properties of VLE related phenomena, such as liquid droplet nucleation from a supersaturated vapor phase \cite{Chen2005}.

\par One notable advanced many-body model for water is the MB-pol potential \cite{Babin2013, Babin2014, Medders2014}. MB-pol was constructed to reproduce interaction energies based on high-level coupled-cluster theory calculations--often referred to as the `gold standard' among quantum chemistry methods--and explicitly accounts for many-body effects through a rigorous many-body expansion of the intra- and intermolecular energies. MB-pol was shown to have excellent agreement with the coupled cluster reference calculations in terms of interaction energies, and accurately reproduces many structural and thermodynamic properties of water, from gas to the condensed phases \cite{Cisneros2016a, Reddy2016, Reddy2017, Moberg2019}. However, the VLE behavior of MB-pol has not yet been comprehensively explored, particularly at elevated temperatures, partially due to the increased computational cost of MB-pol relative to other advanced water models \cite{Reddy2016}.

\par In this work, we used a new implementation of MB-pol in the MBX\cite{mbx} library interfaced with the LAMMPS \cite{Plimpton1995} molecular simulation software. The MBX library serves as an engine to compute potential energies, forces, and virials according to the many-body MB-pol model, and can be in principle interfaced with any desired molecular simulation code (e.g., LAMMPS \cite{Plimpton1995}, i-PI\cite{CERIOTTI2014}, OpenMM\cite{Eastman2013}). Using direct-coexistence MD simulations \cite{Muller2021Guide, Liu2013, Ismail2006, Sega2017, Statt2020, Silmore2017} in this MBX/LAMMPS framework, we assessed MB-pol's ability to reproduce properties such as equilibrium coexistence densities, interfacial tension, and enthalpy of vaporization, for a broad range of elevated temperatures 400 K < $T$ < 600 K. The new coupling of MBX with LAMMPS allowed us to leverage the highly parallelizable nature of the LAMMPS codebase to access the system sizes and time scales necessary to produce converged interfacial simulations. 
\par The article is organized as follows. In section II, we briefly discuss the construction of the MB-pol water model and our new LAMMPS-based implementation, as well as simulation and analysis details. In section III we present a discussion around the results obtained and the overall performance of the potential in predicting water's VLE. Finally, in section IV we provide a summary of the work and mention possible future investigations.

\section{Methods}

\subsection{MB-pol model details and LAMMPS implementation}
MB-pol is a many-body potential energy function (PEF) entirely derived from \textit{ab initio} reference data, and takes advantage of the fast convergence of the many-body expansion (MBE)\cite{mayer1940mayer} of the energy of a system. MBE states that the total energy of a system can be exactly decomposed as 
\begin{equation} \label{eq:mbexpansion} \begin{split}
	E_{N}(1,\dots,N) & = \sum_{i=1}^N V^{\textit{1}B}(i) + \sum_{i<j}^NV^{\textit{2B}}(i,j) \\
	 & + \sum_{i<j<k}^N V^{\textit{3B}}(i,j,k) \\ 
	 & + \dots + V^{\textit{NB}}(1,\dots,N),
	\end{split}
\end{equation}
where $V^{\textit{nB}}(i,...,m)$ defines the n-body energy involving $n$ monomers $i,...,m$ (in this case each `monomer' is one water molecule). MB-pol has an explicit one-body term described by the PES developed by Partridge and Schwenke,\cite{partridge1997determination} and the long range interactions are modeled with two-body dispersion damped with the Tang-Toennies damping function ($V^{\textit{2B}}_{\text{disp}}$)\cite{tang1984improved} and a slightly modified Thole-type model as introduced in the TTM4-F model\cite{burnham2008vibrational} for the electrostatic interactions, which include two-body permanent electrostatics ($V^{\textit{2B}}_{\text{perm}}$) and many-body classical polarization ($V^{\textit{nB}}_{\text{ind}}$). The two- and three-body terms are supplemented with a correction through permutationally invariant polynomials,\cite{braams2009permutationally} $V^{\textit{2B}}_{\text{poly}}$ and $V^{\textit{3B}}_{\text{poly}}$, respectively, fitted to reproduce two- and three-body reference energies calculated with coupled-cluster with single, double, and perturbative triple excitations (CCSD(T)). These polynomial functions correct for the deficiencies of the classical expressions at short range due to quantum effects such as charge penetration, charge transfer, and Pauli repulsion. Further details of the MB-pol PEF can be found the the MB-pol paper series.\cite{Babin2013,Babin2014}

\par In the standard MB-pol model, intramolecular degrees of freedom are unconstrained, necessitating a small timestep size in MD simulations to accurately capture the motion of the hydrogen atoms. In order to access longer simulation time scales, we also tested a rigid variant of the MB-pol model, in which we constrained the H-O bond length and H-O-H angle to the average values obtained from the flexible MB-pol simulations: 0.969 ${\si{\angstrom}}$ and 104.7$^{\circ}$, respectively, via the LAMMPS "fix\_rigid" command. This rigid variant of the model allowed us to increase the timestep by a factor of 10 as detailed below.

The MBX library implementing the MB-pol model\cite{mbx} is tightly coupled with the LAMMPS code\cite{Plimpton1995} with the objective of targeting large-scale molecular simulations on high-performance computing resources. Given the highly modular nature of LAMMPS, the interface to MBX was implemented via new "fix\_mbx" and "pair\_mbx" styles with no modifications to the original LAMMPS code. The MBX library was extended to support the spatial domain decomposition algorithm in LAMMPS to parallelize the computation of energies, forces, and virials. 

\subsection{Simulation and analysis details}

\par We performed all MD simulations with the LAMMPS molecular simulation software \cite{Plimpton1995} interfaced with the MBX library \cite{mbx}. We used a velocity Verlet integration algorithm with a timestep size of 0.2 fs for the flexible model and 2.0 fs for the rigid model. We performed all simulations in the canonical ensemble with the Nos\'e-Hoover thermostat \cite{Kamberaj2005} for temperature control, with LAMMPS thermostat damping parameter 20 fs (flexible) or 200 fs (rigid). The short-range interactions were evaluated with a real-space cutoff of 9.0 ${\si{\angstrom}}$, and long-range interactions (including electrostatic and dispersion contributions) were calculated in reciprocal space using Ewald summation. A detailed list of the standard MBX library parameters used for water\cite{Babin2013, Babin2014, Medders2014, Reddy2016} can be found in the mbx.json files provided in our associated data repository \cite{dataset2021}. To generate the vapor-liquid interfacial system in the direct coexistence geometry, we first equilibrated a liquid configuration in a cubic periodic simulation box, then rapidly expanded the z-dimension of the simulation box to produce a liquid slab in the x-y plane, with the z-dimension normal to the vapor-liquid interface. We then equilibrated the system at the desired temperature for at least 200 ps in all cases. For both rigid and flexible MB-pol variants, we considered a system of 512 molecules in a periodic box of dimensions 20${\si{\angstrom}{\times}}$20${\si{\angstrom}{\times}}$100${\si{\angstrom}}$. To evaluate system size effects we also considered a rigid system of 1024 molecules in a 25.2${\si{\angstrom}{\times}}$25.2${\si{\angstrom}{\times}}$126.0${\si{\angstrom}}$ periodic box. After equilibration, we sampled the flexible systems for 1.1 to 1.5 ns and the rigid systems for 1.5 to 6.0 ns depending on the state point, saving configurations for analysis every 0.2 ps (flexible) or 2 ps (rigid). In the 512 molecule interfacial system, the performance of the flexible MB-pol simulations was $\sim$25 ps/day on 32 2.4-GHz Intel Skylake CPUs, whereas the rigid simulations ran at $\sim$250 ps/day on the same hardware. We note that these performance numbers may change significantly in the future with further improvements to the LAMMPS/MBX implementation. 

\par We calculated equilibrium densities in the vapor phase ($\rho_v$) by computing the average density in the vapor region of the simulation box only considering regions that were at least 10 ${\si{\angstrom}}$ away from the vapor-liquid interface. We then computed the average density profile in the z-dimension and obtained the liquid densities ($\rho_l$) by fitting the density profile with the following hyperbolic tangent-based expression:

    \begin{equation}
        \rho(z) = \frac{\rho_l+\rho_v}{2} - \frac{\rho_l-\rho_v}{2} \mathrm{tanh} \bigg(\frac{z-z_0}{d}\bigg) \label{eq:1}
    \end{equation}
    
Through Eq. \ref{eq:1} we also obtained values for the position of the interface ($z_0$), and the interfacial thickness ($d$). 

\par We estimated the critical parameters $(\rho_c, T_c)$ for each system by fitting the coexistence densities using the following equations\cite{Statt2020}:

  
      \begin{equation}
        \frac{\rho_l+\rho_v}{2} = \rho_c +A(T_c-T) \label{eq:2}
    \end{equation}
    \begin{equation}
           \rho_l-\rho_v = \Delta \rho_0(1-T/T_c)^{\beta} \label{eq:3}
     \end{equation}   
where $A$ and $\Delta \rho_0$ are system-specific parameters to be adjusted in the fitting and the three dimensional Ising model critical exponent $\beta$ is approximately 0.326\cite{ZINNJUSTIN2001159}. Only coexistence densities close to the critical point, from $T$ = 475 K to $T$ = 600 K, were used in this calculation.

\par We calculated surface tension results at each temperature directly from the components of the pressure tensor considering all atoms in the direct coexistence simulations, following the definition:

    \begin{equation}
        \gamma = \frac{L_z }{2}[\langle P_{zz} \rangle -0.5(\langle P_{xx} \rangle+ \langle P_{yy} \rangle)]
    \end{equation}
    
Where $L_z$ is the box length in the z-dimension and $P_{ii}$ are the diagonal components of the pressure tensor. 


\par We computed the enthalpy of vaporization as the enthalpy difference between the gas and liquid states:

    \begin{equation}
        \Delta H_{vap}=H_{gas}-H_{liq}=U_{gas}-U_{liq}+P(V_{gas}-V_{liq}) \label{eq:5}
    \end{equation}
   
where $H$, $U$ and $V$ are the molar enthalpy, molar internal energy and molar volume, respectively, and the subscript indicates if the system is in the liquid or gas state. We obtained the necessary quantities for Eq. \ref{eq:5} by performing separate canonical simulations at the liquid and vapor coexistence densities computed from the direct coexistence simulations. We used a cubic periodic box with 64 molecules for the vapor simulations and 512 molecules for the liquid simulations, and ran these simulations for 200 ps to 2000 ps depending on the state point. All other simulation details were identical to the direct coexistence simulations described above. While in principle, the pressure value used in Eq. \ref{eq:5} could be obtained from the pure liquid, pure vapor, or the direct coexistence simulations (via the z-component of the pressure tensor), we used the pressure obtained from the canonical vapor phase simulations due to their lower uncertainties. We also used this same approach to define the saturation vapor pressure, $P_{vap}$. 
    
\par In all figures and in the SI tables, uncertainties reported are 95\% confidence intervals. For the sake of clarity, statistical uncertainties are shown in figures only when exceeding symbol size, but all uncertainties are reported in the SI. For the coexistence densities, interfacial tension, and enthalpy of vaporization we estimated uncertainties through the standard deviation of 10 block averages. For the critical parameters, we used a bootstrapping approach, similar to what was reported in Ref.~\citen{Statt2020}, in which we fit Eqs. \ref{eq:2} and \ref{eq:3} 300 times with randomly selected coexistence density values obtained from one of the 10 blocks, and estimated 95\% confidence intervals through the standard error of the bootstrap samples.

\section{Results and discussion}

\begin{figure}[!htb]
\centering
\includegraphics[width=1\linewidth]{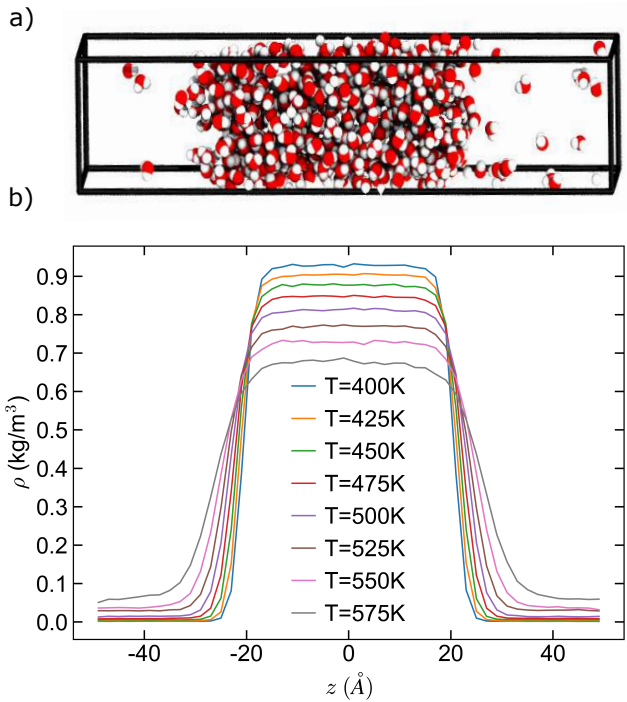}
\caption{(a.) Representative snapshot showing the direct coexistence geometry with 512 molecules at $T$ = 500 K. (b.) Average density profiles obtained from the direct coexistence MD simulations for the MB-pol rigid version with 512 molecules. Colors denote different temperatures as noted.}
\label{fig:profile}
\end{figure}

\begin{figure*}[t]
\centering
\includegraphics[width=1.0\linewidth]{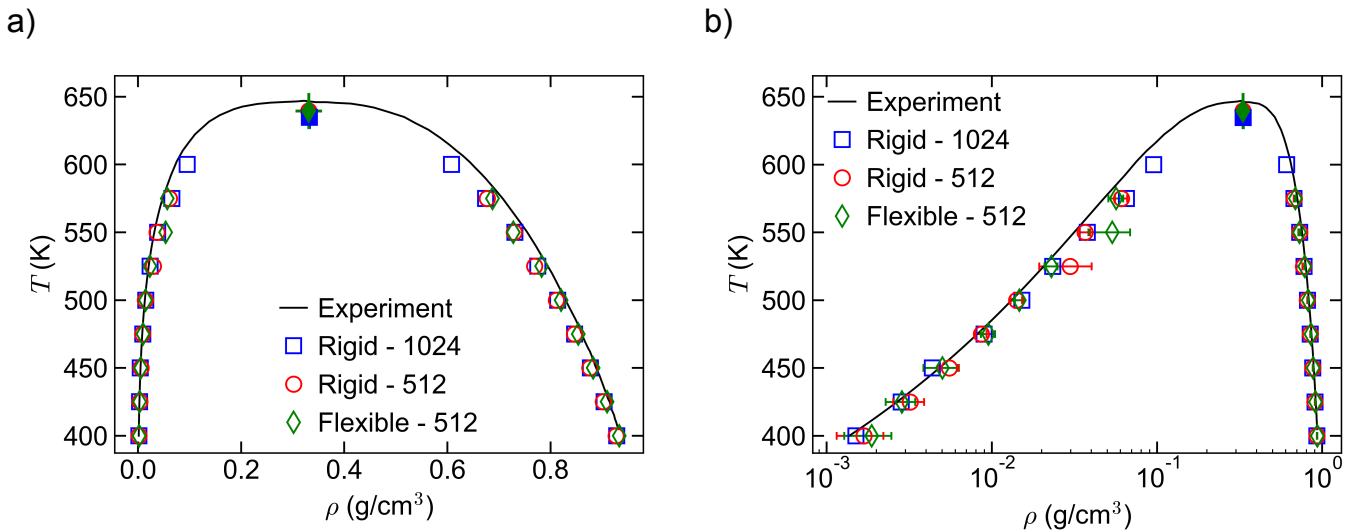}
\caption{(a.) Vapor-liquid coexistence curves (vapor and liquid mass density $\rho$ as a function of temperature $T$) for the three different cases studied: MB-pol rigid model with 512 molecules (red circles), MB-pol rigid model with 1024 molecules (blue squares), MB-pol flexible model with 512 molecules (green diamonds).(b.) Same data as (a.) presented on a log-scale in $\rho$ to display the vapor phase densities. In both panels, coexistence densities are represented as open symbols, estimated critical points are represented as filled symbols, and experimental data \cite{Linstrom} are shown as solid lines. Statistical uncertainties are at the 95\% confidence level and are shown only when larger than the symbol size. Tabulated simulation densities and uncertainties can be found in Supplementary Material.}
\label{fig:vle}
\end{figure*}

\par We begin by assessing the accuracy of the MB-pol model in reproducing VLE coexistence densities of water through comparison with experimental data \cite{Linstrom}. In Figure~\ref{fig:profile} we show a representative simulation snapshot of our direct coexistence geometry at $T$ = 500 K, as well as averaged density profiles for the 512 molecule rigid MB-pol system. As expected, the system maintained a clear separation between liquid and vapor regions, with the vapor density and interfacial width increasing as temperature increased. For the 512 molecule system size, we explored state points ranging from $T$ = 400 K to $T$ = 575 K in order to maintain the reliability of the coexistence density calculations. $T$ < 400 K resulted in too few molecules in the vapor phase for a reliable density estimate, while $T$ > 575 K resulted in poorly-defined liquid and vapor regions. For the larger 1024 molecule system, we added $T$ = 600 K, as the vapor-liquid interface was more stable close to the critical point in the larger system. 
\par Figure~\ref{fig:vle} shows that for all three systems considered, MB-pol vapor and liquid densities were in excellent agreement with experimental results. While the overall agreement in VLE densities was quite good, the phase envelope from MB-pol was predicted to be slightly narrower than the experiments for $T$ > 500 K. The liquid densities of the rigid MB-pol model were slightly lower than the fully flexible version, but the predicted vapor phase densities were within estimated uncertainties in all three cases. The rigid and flexible model variants also displayed only minor differences in liquid structure (Supplementary Material Figure 1). The rigid version showed slightly stronger intermolecular correlations than the flexible version at the same conditions. But, these differences in liquid structure were not enough to significantly affect the VLE density results. There were minimal differences in the coexistence densities from the rigid MB-pol model at different system size.
\par From the coexistence densities we also estimated the critical parameters for each system. For the flexible MB-pol model we found $T_c$ = 639$\pm$14 K and $\rho_c$ = 0.34$\pm$0.03 g/cm$^3$, for rigid MB-pol at 512 molecules $T_c$ = 640$\pm$6 K and $\rho_c$ = 0.33$\pm$0.01 g/cm$^3$, and for rigid MB-pol 1024 molecules $T_c$ = 635$\pm$9 K and $\rho_c$ =0.33$\pm$0.02 g/cm$^3$. These estimated values for the critical parameters were close to the experimental values ($T_c$ = 647.1 K and $\rho_c$ = 0.322 g/cm$^3$) \cite{Linstrom}, and all within uncertainty of one another. The predicted $T_c$ for the 1024 molecule system size was slightly lower than the smaller systems, likely due to the inclusion of one more point ($T$ = 600 K) in the critical point calculation. Relative to experiments, the MB-pol model slightly underestimates the critical temperature and overestimates the critical density. Finite size effects may play a role in the predicted coexistence densities and estimated critical parameters obtained herein, as the limited system dimensions limit the wavelength of density fluctuations that can develop (as well as limiting development of capillary waves along the liquid-vapor interface). However, the impact of the finite system size is nontrivial in an interfacial system \cite{BINDER2000, Longford2018}, and it is difficult to say \textit{a priori} whether in the infinite system our estimated critical parameters would increase or decrease. As such, a rigorous investigation of system size effects--including an extrapolation to the thermodynamic limit--is necessary to fully evaluate these predictions, which is currently too computationally expensive with this model. In future work (as the highly-parallel LAMMPS MB-pol implementation continues to be optimized), we plan to return to the issue of system size effects for a rigorous critical point estimate.


\par
\begin{figure*}
\centering
\includegraphics[width=1\linewidth]{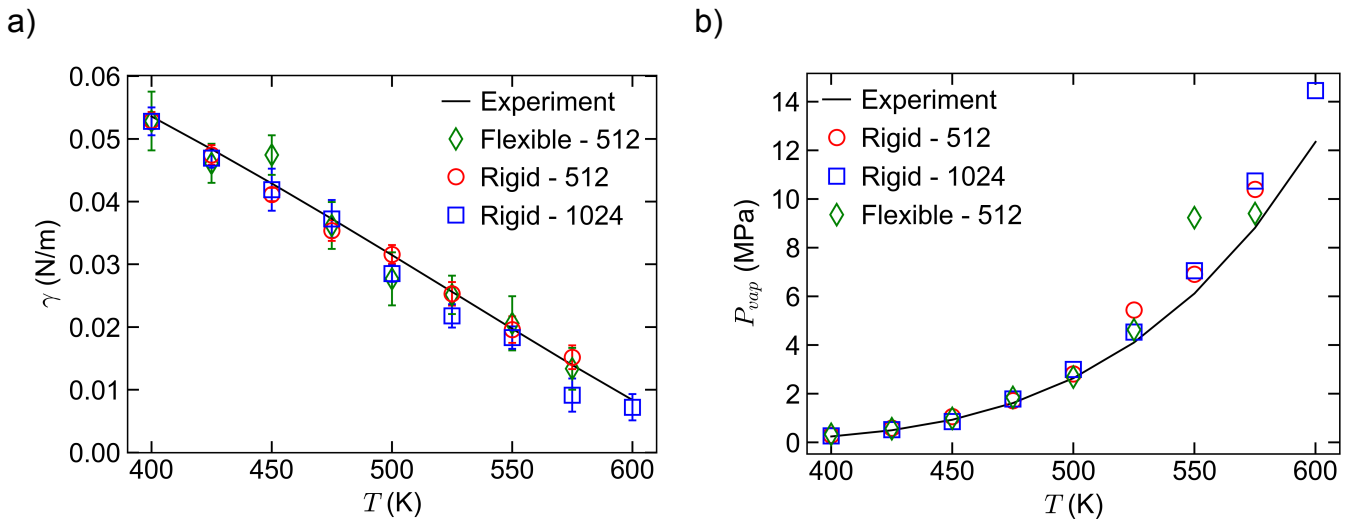}
\caption{(a.) Surface tension, $\gamma$, and (b.) saturation vapor pressure, $P_{vap}$, for rigid MB-pol with 512 molecules (red circles), rigid MB-pol with 1024 molecules (blue squares), flexible MB-pol with 512 molecules (green diamonds) as a function of temperature. Experimental data \cite{Linstrom} are shown as solid lines. Statistical uncertainties are at the 95\% confidence level and are shown only when larger than the symbol size. Tabulated averages and uncertainties can be found in Supplementary Material. }
\label{fig:surface-new}
\end{figure*}

 A key advantage of the direct coexistence approach to VLE calculations is that, in addition to vapor and liquid coexistence densities, one can obtain important characterizations of the interfacial properties. In particular, analysis of the pressure tensor from the simulation results provides access to the vapor-liquid surface tension, $\gamma$ (related to the stability of the interface), as well as the saturation vapor pressure, $P_{vap}$, which we plot in Figure~\ref{fig:surface-new}. For the three systems studied, surface tension results were in excellent agreement with experimental results across the temperature range considered, and did not differ as a function of bond/angle flexibility or system size. Similarly, the $P_{vap}$ agreed closely with the experimental results for $T$ < 500 K and slightly overpredicted the experimental value at high temperatures, which corresponds with the slightly overpredicted vapor density as the system approaches the critical temperature (Figure~\ref{fig:vle}). These quantities provide additional illustrations of the ability of the MB-pol model to accurately reproduce VLE and interfacial properties of water. 


\par
\begin{figure}[!htb]
\centering
\includegraphics[width=1\linewidth]{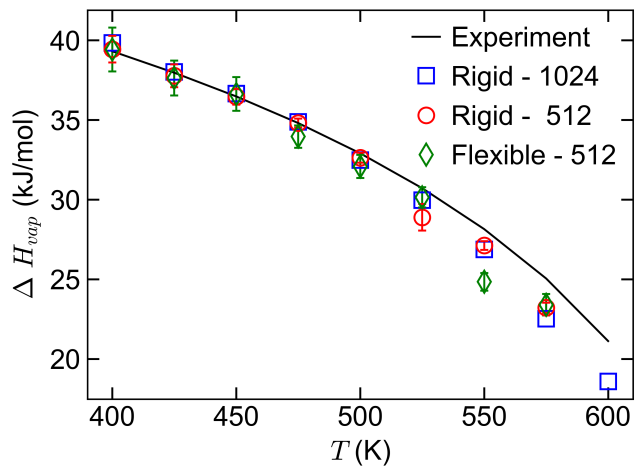}
\caption{Enthalpy of vaporization, $\Delta H_{vap}$, for rigid MB-pol with 512 molecules (red circles), rigid MB-pol with 1024 molecules (blue squares), flexible MB-pol with 512 molecules (green diamonds) as a function of temperature. Experimental data \cite{Linstrom} are shown as solid lines. Statistical uncertainties are at the 95\% confidence level and are shown only when larger than the symbol size. Tabulated averages and uncertainties can be found in Supplementary Material.}
\label{fig:hvap-new}
\end{figure}

 Finally, we also estimated thermodynamic quantities such as the enthalpy of vaporization, $\Delta H_{vap}$, as a function of temperature (Figure~\ref{fig:hvap-new}). Similar to the saturation vapor pressure, $\Delta H_{vap}$ values had no significant differences among the model variants, and agreed closely with experimental data for $T$ < 500 K. Uncertainties were somewhat higher with the flexible MB-pol model, likely due to the shorter accessible simulation timescales. At high temperature, MB-pol slightly underpredicted $\Delta H_{vap}$, again in line with the narrower phase envelope in that regime of temperatures. We note that in Ref. ~\citen{Reddy2016}, which explored water's thermophysical properties for $T$ < 368 K, MB-pol was found to overpredict $\Delta H_{vap}$, with the agreement between simulation and experiment improving as $T$ increased. Combined with our results, this trend indicates a crossover between the MB-pol and experimental $\Delta H_{vap}$ somewhere in the vicinity of $T$ = 400 K. Notwithstanding, we emphasize the overall close agreement between simulation and experimental $\Delta H_{vap}$ across a wide range of temperatures.

\par Taken together, these results show the utility of MB-pol as an accurate predictive model of water's vapor-liquid phase behavior across a wide range of elevated temperatures. MB-pol predicts liquid and vapor coexistence densities, interfacial energetics, and thermodynamic phase change data in close agreement with experimental results. Such agreement is notable because MB-pol was parameterized to reproduce the energetics of isolated 2-body and 3-body clusters, and does not include condensed phases or interfacial configurations in the reference data set.\cite{Babin2013, Babin2014, Medders2014} MB-pol's accuracy in reproducing the properties of liquid water has been well established \cite{Reddy2016}, herein we show that the model's construction is also sufficient for high quality descriptions of the vapor-liquid interface and phase behavior. Another interesting result is the relatively minor differences in the properties of the rigid and flexible versions of MB-pol, allowing one to access significantly larger simulation timescales (while maintaining a similar degree of predictive accuracy) for systems where intramolecular degrees of freedom are unimportant. Unlike fixed-charge models, which tend to show differences in the properties of their rigid and flexible variants \cite{Yuet2010, Raabe2011}, MB-pol’s polarizable nature means that the molecule's dipole moment (and thus the magnitude of the electrostatic partial charges) is free to instantaneously respond to the environment, even when the spatial geometry of the charges is fixed. This aspect resulted in minimal differences upon constraining the intramolecular degrees of freedom for the properties explored herein. Lastly, system size effects appear to be minimal for the interfacial geometries explored in this work, though we plan to more thoroughly explore system size effects as our LAMMPS implementation continues to be optimized for computational efficiency. 

\section{Conclusion}

\par In this work, we investigated the vapor-liquid equilibrium properties of water with the many-body MB-pol potential. Through direct coexistence classical MD simulations, we assessed the accuracy of MB-pol in reproducing water's properties, including coexistence densities, surface tension, vapor pressure, and enthalpy of vaporization. We also evaluated system size effects and the impact of rigid vs. flexible model variants for all properties explored. A new implementation of MB-pol as a library in LAMMPS allowed us to access system sizes and simulation length scales never before achieved with the MB-pol model, and produce converged thermodynamic properties in the direct coexistence geometry.

\par The MB-pol results were in excellent agreement with experimental data, providing additional evidence of MB-pol's ability to accurately reproduce water's thermophysical properties across the phase diagram. Specifically in this paper, we show how MB-pol's transferability can be extended to study a heterogeneous system, water's VLE, which has important industrial applications. Furthermore, the rigid variant of the model was nearly identical to the flexible version, allowing access to even longer length and time scales. In future work, we plan to rigorously explore system size effects to allow extrapolation to the thermodynamic limit. Lastly, a similar approach could be used to investigate the structure of interfacial water from first principles, providing a predictive analysis of hydrogen bonding and water's molecular arrangement at the vapor-liquid interface for elevated temperatures.

\section{Supplementary Material}

\par The supplementary materials include liquid structure for the rigid and flexible MB-pol models, tabulated liquid and vapor coexistence densities, surface tension values, vapor pressures, and enthalpy of vaporization, as well as the statistical uncertainties for these quantities. 

\section{Acknowledgments}
The work of M.C.M, T.E.G., S.Y., and A.Z.P. was supported by the “Chemistry in Solution and at Interfaces” (CSI) Center funded by the U.S. Department of Energy Award DE-SC$001934$, with additional support from Award DE-SC$0002128$. M.R. and F.P. were supported by the U.S. Department of Energy, through Award DE-SC0019490. Computational resources were provided by Terascale Infrastructure for Groundbreaking Research in Engineering and Science (TIGRESS) at Princeton University and the National Energy Research Scientific Computing Center (NERSC), a U.S. Department of Energy Office of Science User Facility operated under Contract No. DE-AC02-05CH11231. This research used resources of the Argonne Leadership Computing Facility, which is a U.S. Department of Energy Office of Science User Facility operated under contract DE-AC02-06CH11357.

\section{AIP PUBLISHING DATA SHARING POLICY}

All data related to this work, including LAMMPS input files, mbx.json parameter files, raw simulation trajectory data, analysis scripts, and processed data used to create all figures in the manuscript, are available for download at the Princeton DataSpace repository at \url{https://doi.org/10.34770/nfcx-rb66}    \cite{dataset2021}.  

\section{References}
\bibliography{ref.bib}

\end{document}